\documentclass[journal]{IEEEtran}
\usepackage{amssymb}
\usepackage{amsmath}
\usepackage{dcolumn}
\usepackage{bm}
\usepackage{color}
\usepackage{graphicx}
\usepackage{subfigure} 
\usepackage{tabularx}
\usepackage{array}
\usepackage{amsthm}
\usepackage{cite}
\usepackage[ruled,linesnumbered]{algorithm2e}
\usepackage{tabu}                     
\usepackage{multirow}                 
\usepackage{multicol}                 
\usepackage{multirow}                
\usepackage{float}                    
\usepackage{makecell}                 
\usepackage{booktabs}
\usepackage{threeparttable} 
\usepackage{booktabs}
\usepackage{threeparttable}  
\usepackage{tikz}

\usepackage{etoolbox}
\makeatletter
\patchcmd{\@makecaption}
  {\scshape}
  {}
  {}
  {}
\makeatother

\SetKwInput{KwData}{Input}
\SetKwInput{KwResult}{Output}

\usepackage[colorlinks,
            linkcolor=blue,
            anchorcolor=green,
            citecolor=red
            ]{hyperref}

\newtheorem{theorem}{Theorem}
\newtheorem{corollary}{Corollary}
\newtheorem{lemma}{Lemma}
\newtheorem{remark}{Remark}
\newtheorem{definition}{Definition}
\newtheorem{assumption}{Assumption}
\newtheorem{proposition}{Proposition}
\newtheorem{problem}{Problem}

\renewcommand{\t}{^{\mbox{\tiny\sf T}}}
\newcommand{\bremark}{\begin{remark}
\begin{rm}}
\newcommand{\eremark}{ \end{rm}\hfill \rule{1mm}{2mm}
\end{remark} }
\newcommand{\btheorem}{\begin{theorem} \begin{it}}
\newcommand{\etheorem}{\end{it} \hfill \rule{1mm}{2mm}
\end{theorem} }
\newcommand{\blemma}{\begin{lemma} \begin{it} }
\newcommand{\elemma}{ \end{it} \hfill\rule{1mm}{2mm}
\end{lemma} }
\newcommand{\bcorollary}{\begin{corollary} \begin{it} }
\newcommand{\ecorollary}{ \end{it} \hfill\rule{1mm}{2mm}
\end{corollary} }
\newcommand{\bdefinition}{\begin{definition} }
\newcommand{\edefinition}{ \hfill\rule{1mm}{2mm}
\end{definition} }
\newcommand{\bproposition}{\begin{proposition} }
\newcommand{\eproposition}{\hfill \rule{1mm}{2mm}
\end{proposition} }
\newcommand{\bexample}{\begin{example} \begin{rm}}
\newcommand{\eexample}{ \end{rm} \hfill\rule{1mm}{2mm}
\end{example} }

\newcommand{\bassumption}{\begin{assumption} }
\newcommand{\eassumption}{\hfill \rule{1mm}{2mm}
\end{assumption} }

\newcommand{\balgorithm}{\medskip\begin{algorithm} \rm}
\newcommand{\ealgorithm}{ \hfill \rule{1mm}{2mm}\medskip
\end{algorithm} }

\newcommand{\basm}{\begin{assumption} \begin{rm} }
\newcommand{\easm}{ \end{rm} \hfill\rule{1mm}{2mm}
\end{assumption} }

\begin{document}




\title{Distributed Platoon Control for Safe and Smooth Crossing at an Unsignalized Intersection---Part I: Stability Analysis}


\author{
\IEEEauthorblockN{
Bin-Bin Hu, Shuwei Pei, Saeed Ahmed and Bayu Jayawardhana
}

 \thanks{
 This work was supported by the NXTGEN Hightech program on Autonomous Factory (https://nxtgenhightech.nl/) and the Holland High Tech (TKI HTSM) strategic program PPS-I Flex HighTech under the project number 24PPS173-CABS.
 }
 \thanks{
Bin-Bin Hu, Shuwei Pei, Saeed Ahmed and Bayu Jayawardhana are with the Engineering and Technology Institute Groningen, University of Groningen, 9747 AG Groningen, The Netherlands (e-mail: b.hu@rug.nl, s.pei@rug.nl, s.ahmed@rug.nl, b.jayawardhana@rug.nl).
}
 }
\maketitle

\begin{abstract}
 We propose a distributed platoon controller that guarantees {\it safe} and {\it smooth crossing} of multi-vehicle systems at an unsignalized intersection. To support the stability analysis, we consider a multi-vehicle platoon maneuvering along a {\it closed path} with {\it a single unsignalized intersection}. 
Unlike traditional bi-level decoupled designs on open paths, which primarily focus on intersection control while neglecting its coupling with platoon convergence, and thus hindering rigorous convergence analysis, the {\it closed-path setting} allows us to analyze the asymptotic behavior of the platoon.
The local exponential convergence analysis of the {\it switched closed-loop system}, induced by repeated vehicle switching at the intersection, is presented. Under some mild conditions on the controller gains, we show that {\it safe intersection crossing} is ensured during the transient phase and {\it smooth crossing} is achieved in steady state, where all vehicles travel at a common constant speed while avoiding collisions at the intersection. Finally, the effectiveness of the proposed distributed controller is demonstrated in a high-fidelity CARLA simulator using multiple commercial vehicle models. 
\end{abstract}


\IEEEpeerreviewmaketitle

\vspace{-0.3cm}

\section{introduction}

With advances in connected and autonomous vehicle (CAV) technologies, multi-vehicle coordination has become increasingly important in urban transportation \cite{lin2017autonomous,fax2004information}. Among cooperative scenarios, connected vehicle platooning has attracted significant attention because it enables vehicles to maintain desired inter-vehicle spacing, improving traffic efficiency, energy consumption, and safety \cite{vegamoor2021string,hu2025c2te}. However, urban intersections remain challenging, as vehicles from different directions must share conflict regions safely and efficiently. Frequent deceleration, acceleration, and stopping may disrupt platoon continuity, causing platoon splitting or performance degradation. Therefore, cooperative intersection control (CIC) for multi-vehicle platooning has emerged as a promising approach to improving future intelligent transportation systems \cite{lee2012development}.

Depending on whether traffic signals are used to manage traffic at intersections, existing studies can be divided into two categories, namely, signal-based and unsignalized methods \cite{rostomyan2025central}.


Following the signal-based CIC research line, the seminal work \cite{caudill1976intersection} studied intersection control for a personal rapid transit system. Later, a isolated controller
based on max-plus models was proposed to reduce energy consumption at homogeneous intersections \cite{haddad2010optimal}, while a multi-agent reinforcement learning method was proposed for heterogeneous intersections by modeling intersections as basic agents \cite{zhu2025sharing}. However, these methods require physical traffic signals, leading to high installation and maintenance costs. This motivates unsignalized approaches with virtual infrastructure and fast computation.


Unsignalized methods have emerged as a promising alternative to signal-based control by reducing unnecessary stops through flexible vehicle coordination. Based on how decision-making, computation, and coordination responsibilities are allocated between infrastructure and vehicles, existing methods can be classified into three categories: centralized, decentralized, and distributed approaches \cite{zhou2025unsignalized}:
\textbf{(i)} For centralized methods where vehicles obtain traffic schemes from the intersection unit, a conflict-duration graph-based framework was designed to improve safety and capacity at unsignalized intersections \cite{deng2020conflict}. To reduce computational burden, an abstraction-based supervision  framework was adopted in \cite{ahn2019abstraction}. For mixed traffic, deep reinforcement learning was used to enable connected autonomous vehicles to yield right of way and optimize traffic flow \cite{yan2021courteous}. For safety guarantee and performance, a polling-systems-based coordination control algorithm was proposed in \cite{miculescu2019polling} with rigorous upper bound. 
However, these methods directly transmit acceleration commands to individual vehicles, making the system sensitive to disturbances. To improve robustness, a CIC controller was proposed in \cite{rostomyan2025central}, assigning each CAV a safe reference path and desired speed based on its state, onboard controller response, and origin--destination pair; \textbf{(ii)} For decentralized methods where vehicles distribute tasks across multiple computing nodes, a virtual platooning approach was proposed to map two-dimensional vehicle motions onto a one-dimensional virtual platoon for efficient intersection operation \cite{medina2017cooperative}. To handle inaccuracies and disturbances, a hybrid CIC framework was developed by integrating microscopic virtual platooning with traffic-flow regulation \cite{zhou2022cooperative}. Later, a constraint-coupled bi-level optimization method was designed to balance performance and computational efficiency \cite{deng2023cooperative}.

In contrast to the centralized and decentralized methods ~\cite{deng2020conflict,ahn2019abstraction,yan2021courteous,miculescu2019polling,medina2017cooperative,zhou2022cooperative,deng2023cooperative} in~\textbf{(i)-(ii)}, the distributed control policies in \textbf{(iii)} have also attracted considerable attention. Along this line, a distributed optimization algorithm was proposed in \cite{bian2019cooperation} to schedule conflict-free arrival times for trajectory planning at intersections. To minimize travel time and speed variations simultaneously, a mixed-integer nonlinear optimization method was developed in~\cite{mirheli2019consensus} to avoid near-crash conditions. Moreover, a graph-based cooperation method was proposed in \cite{chen2021graph} to formulate the conflict-free scheduling problem at unsignalized intersections, which enhances control flexibility.

However, most existing CIC methods \cite{rostomyan2025central,caudill1976intersection,haddad2010optimal,zhu2025sharing,zhou2025unsignalized,deng2020conflict,ahn2019abstraction,yan2021courteous,medina2017cooperative,zhou2022cooperative,deng2023cooperative,bian2019cooperation,mirheli2019consensus,chen2021graph} provide only limited rigorous analysis, mainly due to their bi-level decoupled designs for intersection control and platoon convergence on open paths. To facilitate stability analysis and safety guarantee, we consider a setting in which the platoon maneuvers along a {\it closed path} with {\it a single unsignalized intersection}, enabling the platoon to be analyzed as an integrated team. Compared with open-path scenarios, this closed-path setting introduces two challenges: \textbf{(i)} Vehicles repeatedly enter the intersection, which causes persistent coupling between intersection control and platoon convergence and invalidates decoupled optimization designs; \textbf{(ii)} Repeated intersection entries transform the closed-loop dynamics into a switched system, which makes the convergence analysis challenging.



To tackle these challenges, we 
propose a distributed controller that enables multiple vehicles to cross the intersection safely and converge to the steady-state platoon, where all vehicles travel at a common constant speed and achieve collision-free crossing. 
The local exponential convergence of the {\it switched closed-loop system}, induced by repeated vehicle switching at the intersection, is rigorously analyzed. 
Finally, the effectiveness of the proposed algorithm is demonstrated. 

\noindent \emph{Contribution:} The contribution is summarized as follows.
\begin{enumerate}
\item  
We propose a distributed smooth-crossing controller that enables multiple vehicles to spontaneously form a platoon on a closed path with a single intersection; 

\item
We enable smooth constant-speed crossing without acceleration or deceleration in steady state by introducing a sensing barrier function for incoming vehicle pairs approaching the intersection;

\item
We rigorously establish the local exponential convergence of the switched closed-loop system induced by vehicle switching at the intersection on a closed path.




\end{enumerate}

The proposed algorithm is not intended to replace general-purpose intersection scheduling methods. Rather, it provides a control-theoretic model in which the interaction between intersection safety and platoon convergence remains active indefinitely, which leads to a switched closed-loop system under repeated intersection crossings. This allows the switching dynamics to be included in the rigorous convergence analysis.

\noindent\emph{Notation:} We denote the sets of real, positive real, and integer numbers by $\mathbb{R}$, $\mathbb{R}^+$, and $\mathbb{Z}$, respectively. For $i\leq j \in\mathbb{Z}$, let $\mathbb{Z}_i^j:=\{m\in\mathbb{Z}\mid i\leq m\leq j\}$. For a signal $z\in\mathbb{R}$, its magnitude is denoted by $|z|$. The $n$-dimensional identity matrix and zero vector are denoted by $I_n$ and $\textbf{0}_{n}$, respectively.

\vspace{-0.3cm}
\section{Preliminary}
\label{sec_prelimimary}

\begin{figure}[!htb]
\centering
\includegraphics[width=\hsize]{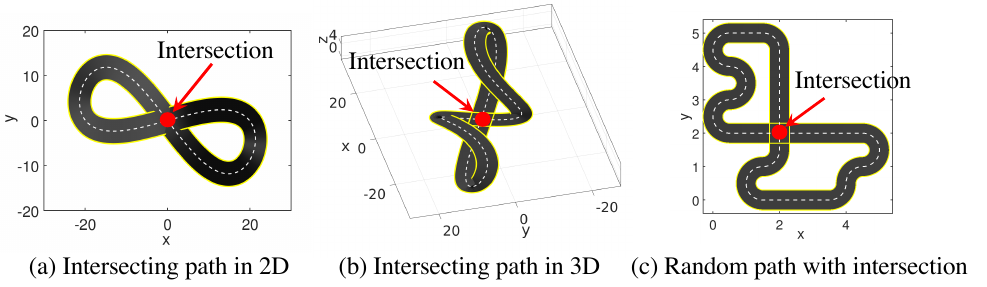}
\caption{ Different types of single-intersection closed paths under the interpretation of \eqref{closed_path} and \eqref{single_intersection}.
(a) The eight-shape path with a single intersection at the center point in 2D. (b) The path with a single intersection at the center point in 3D. (c) Random urban path with a single intersection. }
\label{path_type}
\end{figure}
\subsection{Single-Intersection Closed Paths}
\label{single_intersection_closed_paths}
Suppose that a closed path $\mathcal P$ with a single intersection 
in the 2D plane can be characterized by the following zero-level parametric functions:
\begin{align}
\label{closed_path}
\mathcal P = &\{\bold{p}:=[x,~y]\in\mathbb{R}^2~|~\phi_1(\bold{p}):=x-f_1(\theta)=0,\nonumber\\
&\phi_2(\bold{p}):=y-f_2(\theta)=0,\; \theta\in\mathbb S \},
\end{align}
where $x\in\mathbb{R}$ and $y\in\mathbb{R}$ denote the Cartesian coordinates, $\theta$ represents the phase variable in the phase space $\mathbb{S}=\{\theta~|~\theta\in[0,2\pi)\}$, and the functions $f_1, f_2$ describe the mapping from the phase space $\mathbb S$ to the Euclidean space $\mathbb{R}$.  
The parametric interpretation \eqref{closed_path} has two advantages: First, by choosing different functions $f_1$ and $f_2$, it can describe arbitrary closed road paths with a single intersection, as illustrated in Fig.~\ref{path_type}. Second, it enables vehicle control in the phase coordinate before mapping the results back to physical space, avoiding complicated geodesic-distance computations.

For the single intersection, denoted by $\bold{p}_{s}:=[x_s, y_s]^\top$, we assume {\it without loss of generality} that it is located at the center of the entire path and is shared by two parameter values $\theta=0$ and $\theta=\pi$, i.e.,
\begin{align}
\label{single_intersection}
x_s = f_1(0) = f_1(\pi),~y_s = f_2(0) = f_2(\pi).
\end{align}
As an initial theoretical study of platoon convergence on closed paths with an intersection, we consider a centrally located intersection. This configuration yields approximately balanced numbers of vehicles in $\theta \in [0,\pi)$ and $\theta \in [\pi,2\pi)$ at steady states, which facilitates the switched-system convergence analysis. In contrast, a noncentral intersection generally leads to uneven vehicle distributions and more complicated switching sequences, requiring additional analytical tools rather than a direct extension of the present results. Such cases are therefore left for future work.

\begin{remark}
The study of single-intersection closed paths remains both significant and meaningful, as it captures cyclic urban traffic scenarios, such as local ring roads and other closed-loop road structures, where vehicles may repeatedly encounter the same intersection \cite{hmdaORR}. Unlike open-path settings, the vehicles on a closed path affect those approaching the intersection after crossing it. This repeated crossing gives rise to a challenging switched system and motivates the analysis developed in this work.
\end{remark}





\begin{figure}[!htb]
\centering
\includegraphics[width=8.5cm]{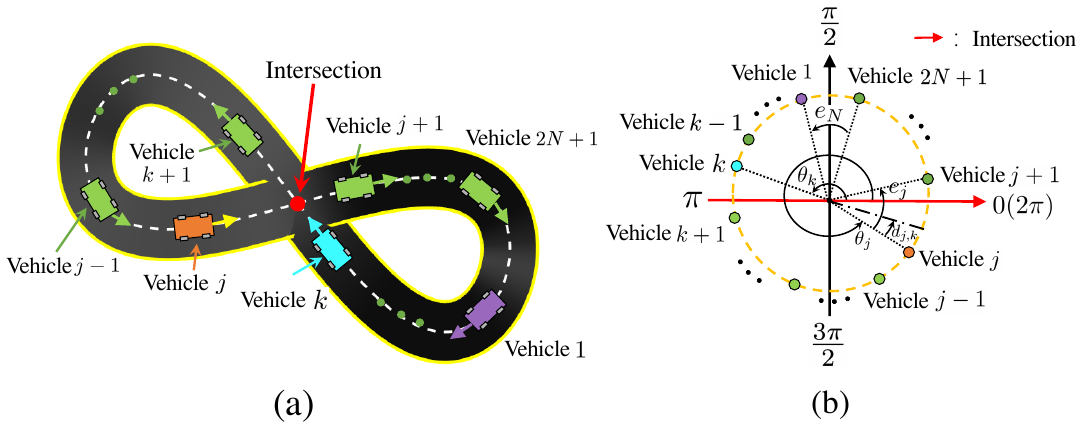}
\caption{(a) Illustration of the vehicle platoon on a figure-eight road with an intersection in Fig.~\ref{path_type}. (b) The corresponding displacements in the phase space. The brown and blue vehicles in (a) represent vehicles that are ready to enter the intersection. The red point denotes the intersection.} 
\label{vehicle_frame}
\end{figure}
\subsection{Multi-Vehicle Systems}
As shown in Fig.~\ref{vehicle_frame}, based on the parametric path $\mathcal P$ in \eqref{closed_path}, we consider a team of vehicles indexed by $\mathcal V=\{1,2,\dots,2N+1\}$ maneuvering on this one-dimensional path. 
Consequently, the dynamics of each vehicle $i$ can be characterized by a phase variable $\theta_i$ as given below
\begin{align}
\label{phase_dynamics}
\dot{\theta}_i = u_i, \quad i \in \mathcal V,
\end{align}
where $\theta_i \in [0, 2\pi)$ denotes the phase position of vehicle $i$ on~the path $\mathcal P$, and $u_i \in \mathbb{R}$ represents the control input of vehicle $i$ in the phase space. For more complicated vehicle dynamics, we can leverage a hierarchical control framework by regarding the input $u_i$ in \eqref{phase_dynamics} as a high-level command \cite{hu2025c2te}.


\subsection{Safe- and Smooth-Crossing Platoon}
\label{sub_sec_smooth}
To achieve safe and smooth platoon crossing, we define the phase-spacing error between adjacent vehicles $i$ and $i+1$ as
\begin{equation}
\label{relative_angle}
e_i
=
\theta_{i+1}-\theta_i+\xi_i
-\frac{2\pi}{2N+1},
\end{equation}
where cyclic indexing is adopted, i.e., $\theta_{2N+2}=\theta_1$,
$\theta_i$ is defined in \eqref{phase_dynamics},
and $|\mathcal V|=2N+1$ denotes the number of vehicles.
The variable $\xi_i$ is defined as
\begin{equation}
\label{BIK_GVF_carrier}
\xi_i=
\left\{
\begin{aligned}
0, \qquad &\mathrm{if}~\theta_{i+1}-\theta_i\geq 0,\\
2\pi, \qquad &\mathrm{if}~\theta_{i+1}-\theta_i<0,
\end{aligned}
\right.
\end{equation}
which accounts for the wrap-around effect and ensures that
$\theta_{i+1}-\theta_i+\xi_i$ represents the forward cyclic distance.

To prevent potential collisions of vehicles at the intersection in Fig.~\ref{vehicle_frame} 
(a), 
we also define two vehicles $j$ and~$k$ that are about to enter the intersection as follows:
\begin{align}
\label{two_vehicle_intersection_1}
j=\arg\max_{\substack{i\in\mathcal V}}\{ \theta_i~|~\theta_i<2\pi\}, k=\arg\max_{\substack{i\in\mathcal V}}\{ \theta_i~|~\theta_i<\pi\}.
\end{align}
Since we will focus on the local stability near the desired platoon, both candidate sets $j$ and $k$ in 
\eqref{two_vehicle_intersection_1} are considered nonempty and well-defined.
Then, it follows from \eqref{two_vehicle_intersection_1} that the relative phase distance $d_{j,k}$ between vehicles $j, k$ becomes
\begin{align}
\label{distance_two_vehicle}
d_{j,k}=2\pi-\theta_{j}-(\pi-\theta_{k})=\theta_{k}-\theta_{j}+\pi.
\end{align}

\begin{definition}
\label{def_platoon_avoidance}(Safe- and Smooth-Crossing Platoon)
For the single-intersection closed path $\mathcal P$ in \eqref{closed_path}, all the vehicles~$\mathcal V$ governed by \eqref{phase_dynamics} collectively achieve smooth platoon maneuvering along the path and through the intersection, if the following four objectives are fulfilled, 
\begin{description}
\item[{\bf P1.}] \label{platoon_obj_1} {\bf (Platoon convergence):} The vehicle platoon is achieved by maintaining identical relative phase between adjacent vehicles, i.e., 
$\lim_{t\rightarrow\infty}e_i(t)=0, i\in\mathcal V$,
where $e_i$ is given in \eqref{relative_angle}.



\item[{\bf P2.}] \label{platoon_obj_2} {\bf (Adjacent-vehicle collision avoidance):}  The collision avoidance between adjacent vehicles is guaranteed, i.e., $\theta_{i+1}(t)-\theta_i (t)+\xi_i(t)> 0$ for all $ i\in\mathcal V,  t\geq0$ with $\xi_i$ given in \eqref{relative_angle}.

\item[{\bf P3.}] \label{platoon_obj_3} {\bf (Smooth crossing at the intersection):} All the vehicles maintain a desired angular velocity along the path without acceleration or deceleration, i.e., $\lim_{t\rightarrow\infty}\dot{\theta}_i(t)=\lim_{t\rightarrow\infty}\dot{\theta}_l(t)=v_d$ for all $i\neq l\in\mathcal V$, 
where $v_d\in\mathbb{R}^+$ is a predetermined velocity.

\item[{\bf P4.}] \label{platoon_obj_4} {\bf (Collision avoidance at the intersection):} Two vehicles $j$ and $k$ in \eqref{two_vehicle_intersection_1} that are about to enter the intersection are guaranteed to avoid collisions, i.e.,
$|d_{j,k}(t)|\neq 0$, for all $ t>0$ with $d_{j,k}$ given in \eqref{distance_two_vehicle}.

\end{description}
\end{definition}

Unlike conventional platoon strategies \cite{vegamoor2021string,hu2025c2te,medina2017cooperative,zhou2022cooperative,deng2023cooperative} which primarily focus on objectives {\bf P1–P2}, the presence of a single intersection along the central path inevitably introduces persistent vehicle interactions at the intersection, which can significantly disrupt the smooth maneuvering of the platoon and lead to additional requirements of {\bf P3–P4} in Definition~\ref{def_platoon_avoidance}. 





\subsection{Problem Formulation}

%


\begin{problem}
\label{problem_deadlock}
Design the distributed control input $u_i$ for each vehicle $i$ in \eqref{phase_dynamics} such that the {\it safe-} and {\it smooth-crossing platoon} in Definition~\ref{def_platoon_avoidance} is achieved.
\end{problem}


\begin{assumption}
\label{assump_two_vehicle_initial_position}
The initial relative phase between any two adjacent vehicles satisfies $\theta_{i+1}(0)-\theta_{i}(0)+\xi_i(0) > 0$ for all $i\in\mathcal V$ with $\xi_i$ given in \eqref{relative_angle}.
\end{assumption}

\begin{assumption}
\label{assump_two_vehicle_intersection}
The initial relative phase distance between two vehicles at the intersection $d_{j,k}$ in \eqref{distance_two_vehicle} satisfies $|d_{j,k}(0)|>0$.
\end{assumption}

Assumptions~\ref{assump_two_vehicle_initial_position} and \ref{assump_two_vehicle_intersection} specify the initial conditions needed to avoid vehicle collisions at the intersection.



\section{Controller Design and Error Dynamics}
\label{sec_controller}


\subsection{Equilibrium Existence }
\label{existence_platoon}
Before presenting the controller design, we first establish the existence of the {\it safe-} and {\it smooth-crossing platoon} satisfying {\bf P1-P4} in Definition~\ref{def_platoon_avoidance}.

\begin{lemma}
\label{lemma_existence}
For a team of $2N+1$ vehicles governed by \eqref{phase_dynamics} and the single-intersection closed path $\mathcal{P}$ given in \eqref{closed_path} and~\eqref{single_intersection}, suppose that the steady platoon (i.e., equilibrium) is achieved on the path $ \mathcal {P} $; there always exists a {\it safe-} and {\it smooth-crossing platoon} satisfying {\bf P1–P4} such that no more than one vehicle occupies the intersection all along, i.e.,
\begin{equation}
\label{existence_solution}
\left| \mathcal I(t) \right| \le 1,
\quad \forall t \ge T_s,
\end{equation}
with the set $\mathcal I:=\big\{i \in \mathbb{Z}_1^{2N+1}~\big|~\theta_i(t)\in\{0,\pi\} \big\}$ represents the index of vehicles at the intersection, and the time $T_s\in\mathbb{R}^+$ denotes the time at which the steady platoon is achieved.
\end{lemma}

\begin{proof}
We will prove the condition \eqref{existence_solution} by contradiction. Without loss of generality, we assume that there exists at least 
two vehicles labeled $l,m\in\mathbb{Z}_{1}^{2N+1}$ satisfying 
\begin{align}
\label{two_vehicle_intersection}
\theta_l(T_1)=0, \theta_m(T_1)=\pi~\mathrm{such~that}~|\mathcal I|=2 
\end{align}
at the time $T_1>T_s$. Then, it follows from \eqref{two_vehicle_intersection} that $\theta_{m}(T_1)-\theta_l(T_1)=\pi$.
Meanwhile, since the steady platoon has been achieved, it follows from the properties {\bf P1, P3} that the relative phase between arbitrary adjacent vehicles maintains ${2\pi}/{(2N+1)}$ at $t=T_1$, i.e.,
\begin{align}
\label{adjacent_condition}
\theta_{i+1}(T_1)-\theta_{i}(T_1)+\xi_i(T_1)=\frac{2\pi}{2N+1}, \forall i\in\mathcal{V}.
\end{align}
which further implies that $\theta_m(T_1)-\theta_l(T_1)=\theta_m(T_1)-\theta_{m-1}(T_1)+\xi_{m-1}(T_1)\cdots+\theta_{l+1}(T_1)-\theta_l(T_1)+\xi_{l}=\frac{2\pi}{2N+1}a$
with a positive integer $a\in\mathbb{Z}_1^{2N+1}$. However, there is no integer $a$ such that $a{2\pi}/({2N+1})=\pi$, so we arrive at a contradiction. Therefore, we can guarantee the condition \eqref{existence_solution}. The proof is completed.
\end{proof}

\subsection{Distributed Controller \& Switched Error Dynamics}
Since we have shown the existence of a {\it safe-} and {\it smooth-crossing platoon} solution in Section~\ref{existence_platoon}, we can design the distributed {\it safe-} and {\it smooth-crossing platoon} controller for each vehicle $i$ to be
\begin{align}
\label{vehicle_controller}
u_i=v_d+k_1e_i+k_2\eta_i,
\end{align}
where $k_1 \in \mathbb{R}^+$ and $k_2 \in \mathbb{R}^+$ are control gains, the predetermined velocity $v_d$ is given in {\bf P3} of Definition~\ref{def_platoon_avoidance}, the phase error $e_i$ is given in \eqref{relative_angle} and $\eta_i$ is an additional term that accounts for collision avoidance with vehicles $j,k$ in \eqref{two_vehicle_intersection_1} and smooth crossing at the intersection in {\bf P4}, given by
\begin{equation}
\label{potential_function}
\eta_i=\left\{
\begin{aligned}
-\alpha(|d_{j,k}|)\frac{d_{j,k}}{|d_{j,k}|},\qquad&\mathrm{if}~i=j,\\
\alpha(|d_{j,k}|)\frac{d_{j,k}}{|d_{j,k}|},\qquad&\mathrm{if}~i=k,\\
0,\qquad&\mathrm{otherwise}.
\end{aligned}
\right.
\end{equation}
In particular, the barrier function $\alpha(s)$ is given by
\begin{equation}
\label{alpha}
\alpha(s)=\left\{
\begin{aligned}
R-s,\qquad&\mathrm{if}~0<s\leq R,\\
0,\qquad&\mathrm{if}~s>R,
\end{aligned}
\right.
\end{equation}
where $R={\pi}/{(2N+1)}$ is the predetermined sensing radius, and corresponds to half of the desired phase spacing. Note that $\alpha(s)$ is continuous at the switching point $s=R$.

Let the error be $\bold{e}=[e_1, \dots, e_{2N+1}]\t$ in \eqref{relative_angle} and it follows from \eqref{phase_dynamics}, \eqref{vehicle_controller}, \eqref{potential_function} that the derivative of $\bold{e}$ in \eqref{relative_angle} satisfies
\begin{align}
\label{close_loop}
\dot{\bold{e}}=&A\bold{e}+B(t)\alpha(|d_{j,k}|)\frac{d_{j,k}}{|d_{j,k}|},
\end{align}
where the matrices $A, B$ are given by
\begin{align}
\label{error_matrix}
A=&
{\scriptsize\begin{bmatrix}
-k_1 & k_1 & 0 & \cdots & 0\\
0  & -k_1 & k_1 & \dots & 0\\
\vdots & \vdots & \vdots & \ddots & \vdots\\
k_1 & 0 & \cdots & 0 & -k_1\\
\end{bmatrix}}\in\mathbb{R}^{(2N+1)\times (2N+1)},\nonumber\\
\nonumber\\
\begin{tikzpicture}[overlay, remember picture]
    \node[anchor=west] at (1.4, 0.5) {$\scriptstyle (j-1)$-$\scriptstyle\mathrm{th}$};
\end{tikzpicture}
\begin{tikzpicture}[overlay, remember picture]
    \node[anchor=west] at (2.7, 0.5) {$\scriptstyle j$-$\scriptstyle\mathrm{th}$};
\end{tikzpicture}
\begin{tikzpicture}[overlay, remember picture]
    \node[anchor=west] at (3.8, 0.5) {$\scriptstyle (k-1)$-$\scriptstyle\mathrm{th}$};
\end{tikzpicture}
\begin{tikzpicture}[overlay, remember picture]
    \node[anchor=west] at (5.1, 0.5) {$\scriptstyle k$-$\scriptstyle\mathrm{th}$};
\end{tikzpicture}
B=&{\scriptsize\begin{bmatrix}
0 & \cdots & \overbrace{-k_2} & \overbrace{k_2} & \cdots & \overbrace{k_2} & \overbrace{-k_2} \cdots & 0
\end{bmatrix}}\t\in\mathbb{R}^{2N+1}.
\end{align}
As shown in \eqref{error_matrix}, the matrix $B(t)$ varies as different vehicle pairs $j$ and $k$ approach the intersection, which implies that the closed-loop system $\bold{e}$ in \eqref{close_loop} exhibits switching behavior. Since the vehicle passing order is generally unknown, the resulting switching sequence cannot be determined a priori, which complicates the convergence analysis of the switched systems \eqref{close_loop}. To characterize this switching sequence, we consider the case where the vehicle platoon is initialized near the desired configuration and introduce the following assumption.




\begin{assumption}
\label{initial_error}
Given a small positive parameter $\epsilon $ satisfying $
0<k_1\epsilon<k_2R$ and $\sqrt{2N+1}\epsilon<{2\pi}/({2N+1})$,
the initial spacing error in \eqref{relative_angle} satisfies $|e_i(0)| < \epsilon$ for all $i \in \mathbb{Z}_1^{2N+1}$.
\end{assumption}

Assumption~\ref{initial_error} restricts the analysis to a local neighborhood of the desired platoon: the first inequality lets the safety correction dominate the spacing error near the intersection, while the second is used to preserve adjacent-vehicle separation. This local condition is consistent with the closed-path setting considered here, where vehicles repeatedly traverse the same intersection after the platoon has been initialized near its desired cyclic configuration.



\section{Main results}
\label{sec_main_results}





The following theorem presents our main result that shows the controller \eqref{vehicle_controller} under Assumptions~\ref{assump_two_vehicle_initial_position}--\ref{initial_error} solves the {\it safe-} and {\it smooth-crossing platoon} with local exponential convergence.

\begin{theorem}
\label{theorem_main}
Under Assumptions~\ref{assump_two_vehicle_initial_position}-\ref{initial_error}, all the vehicles $\mathcal V$ governed by the controller \eqref{vehicle_controller} and the switching sequence in~\eqref{cycle_sequence} achieve the {\it safe-} and {\it smooth-crossing platoon} with local exponential convergence, i.e., {\bf P1-P4} of Definition~\ref{def_platoon_avoidance}, if the parameters $k_1, k_2$ in \eqref{vehicle_controller} satisfy 
\begin{equation}
\label{convergence_parameter_k}
k_2
<
\frac{(1-\cos\frac{2\pi}{2N+1})k_1-c}{(\sqrt{N+1}-1)}, \ \ k_1>0, k_2>0,
\end{equation}
where $c\in\mathbb{R}^+$ is a sufficiently small positive constant. 
\end{theorem}

Prior to proving the theorem, let us present a number of technical lemmas as follows. Built upon these lemmas, we conclude this section with the proof of Theorem~\ref {theorem_main}. 

 

\subsection{Reduction to Switched Systems with Same Eigenvalues}
\label{sub_reduced}
Although Assumption~\ref{initial_error} restricts the possible switching sequences, it does not rigorously determine the switching order at the intersection. Therefore, we first achieve the collision avoidance at the intersection in {\bf P4}, which allows us to further characterize the switching sequence.




\begin{lemma}
\label{lemma_intersection_avoidance}
Under Assumptions~\ref{assump_two_vehicle_intersection}-\ref{initial_error}, if every pair of vehicles $j$, $k$ satisfies 
$|d_{j,k}(0)|>0$, then the controller \eqref{vehicle_controller}- \eqref{potential_function} achieves the collision avoidance at the intersection, i.e., $|d_{j,k}(t)|>0$ for all $t>0$.
\end{lemma}

\begin{proof}
Taking the derivative of $d_{j,k}$ in \eqref{distance_two_vehicle} and substituting the controller in \eqref{vehicle_controller} and \eqref{potential_function} yields
\begin{equation}
\label{derivative_d_jk}
\dot{d}_{j,k}=\dot{\theta}_{k}-\dot{\theta}_j=k_1(e_k-e_j)+2k_2\alpha(|d_{j,k}|)\frac{d_{j,k}}{|d_{j,k}|}.
\end{equation}
According to $\alpha$ in \eqref{alpha} and Assumption~\ref{assump_two_vehicle_intersection}, we divide the analysis into two cases based on initial value of $|d_{j,k}(0)|$.


{\bf Case 1}: $0<|d_{j,k}(0)|\leq R$. If Case 1 holds, it follows from \eqref{alpha} that the variable $\dot{d}_{j,k}$ becomes
\begin{align}
\label{derivative_d_jk_1}
\dot{d}_{j,k}=k_1(e_k-e_j)+2k_2(R-|d_{j,k}|)\frac{d_{j,k}}{|d_{j,k}|}.
\end{align}
Let $s_{j,k}=R-|d_{j,k}|$, it follows from \eqref{derivative_d_jk_1} that the derivative of $s_{j,k}$ becomes
\begin{align}
\label{derivative_s_jk}
\dot{s}_{j,k}=&~-\frac{d_{j,k}}{|d_{j,k}|}k_1(e_k-e_j)-2k_2s_{j,k}.
\end{align}
Together with Assumption~\ref{initial_error}, this yields $\dot{s}_{j,k}$ in \eqref{derivative_s_jk} becomes
\begin{align}
\label{derivative_s_jk_1}
\dot{s}_{j,k}\leq&~-2k_2 s_{j,k}+2k_1\epsilon.
\end{align}
Using the comparison principle~\cite{khalil2002nonlinear}, one has that
\begin{align}
\label{evolution_s_jK}
s_{j,k}(t)\leq&\frac{k_1}{k_2}\epsilon+\bigg(s_{j,k}(0)-\frac{k_1}{k_2}\epsilon\bigg)e^{-2k_2t}.
\end{align}
Since $s_{j,k}(0)=R-|d_{j,k}(0)|< R$ in Case 1, and ${k_1}\epsilon/{k_2}<R$ in Assumption~\ref{initial_error}, the comparison bound in \eqref{evolution_s_jK} remains below $R$ for all $t>0$. Hence, $s_{j,k}(t)<R$ for all $t>0$. According to the definition of $s_{j,k}=R-|d_{j,k}|$, we obtain that $|d_{j,k}(t)|>0$ for all $t>0$. The proof for Case~1 is thus completed.

{\bf Case 2}: $|d_{j,k}(0)|>R$. If Case 2 holds, the collision avoidance between vehicles $j$ and $k$ at the intersection is initially satisfied, i.e., $|d_{j,k}(0)|\neq 0$. If $|d_{j,k}(t)|>R$ for all $t>0$, then $|d_{j,k}(t)|\neq 0$ is guaranteed. Otherwise, if $|d_{j,k}(t)|$ enters the region of $|d_{j,k}(t)|\le R$ at some time $T_2$, the analysis becomes Case 1. The proof for Case 2 is completed.
\end{proof}

\begin{figure}[!htb]
\centering
\includegraphics[width=\hsize]{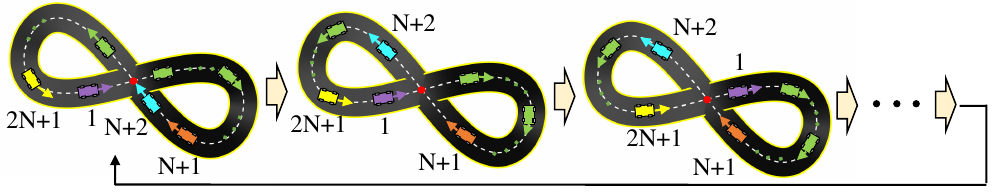}
\caption{Illustration of the periodic switching sequences in~\eqref{cycle_sequence}, where vehicles from the two sides of the intersection pass through alternately.} 
\label{ordering_illustration}
\end{figure}

Based on Lemma~\ref{lemma_intersection_avoidance}, if all vehicles start in a neighborhood of the desired platoon under Assumptions~\ref{assump_two_vehicle_intersection}--\ref{initial_error}, the intersection switching behavior between vehicles $j$ and $k$ can be characterized as shown in Fig.~\ref{vehicle_frame} (a). Without loss of generality, the switching sequence of the pair $(j,k)$ follows the finite cycle:
\begin{align}
\label{cycle_sequence}
(j,k):=&(1,N+2)\rightarrow (1,N+1)\rightarrow (2N+1,N+1)\nonumber\\
&\rightarrow(2N+1,N)\rightarrow\cdots\rightarrow (1,N+2).
\end{align}
As shown in Fig.~\ref{ordering_illustration}, the finite switching sequence in \eqref{cycle_sequence} induces periodic switching in the error dynamics of $\dot{\bold{e}}$ in \eqref{close_loop}, with a fixed switching order. However, the error dynamics still contain additional terms of $\alpha(|d_{j,k}|)d_{j,k}/|d_{j,k}|$, which are not explicitly expressed in terms of $\bold{e}$ and thus complicate the analysis. 
In what follows, we first express these terms in terms of $\bold{e}$ for a representative switching mode and then extend the formulation to all modes along the switching sequence.



In particular, consider the switching mode in which the $j$-th and $k$-th vehicles in Fig.~\ref{vehicle_frame} are indexed as the first and $(N+2)$-th vehicles, respectively, i.e.,
\begin{align}
\label{initial_intersection}
j=1,\quad k=N+2.
\end{align}
Suppose that the phase distances of the two vehicles from the intersection satisfy $2\pi-\theta_1 > \pi-\theta_{N+2}$.
Recalling \eqref{relative_angle} and \eqref{distance_two_vehicle}, the relative phase distance $d_{1,N+2}$ becomes
\begin{align}
d_{1,N+2}=&\pi-\theta_{1}+\theta_{N+2}\nonumber\\
=&\pi+(\theta_2-\theta_1+2\pi-\frac{2\pi}{2N+1}+\theta_3-\theta_2\nonumber\\
&-\frac{2\pi}{2N+1}+\cdots+\theta_{N+2}-\theta_{N+1}-\frac{2\pi}{2N+1})\nonumber\\ 
&+\frac{2\pi(N+1)}{2N+1}-2\pi\nonumber\\
=&\sum_{i=1}^{N+1} e_i+\frac{\pi}{2N+1}>0.
\end{align}
By choosing $R={\pi}/({2N+1})$, it follows from \eqref{alpha}
that 
\begin{align}
\label{new_alpha}
\alpha(|d_{1,N+2}|)\frac{d_{1,N+2}}{|d_{1,N+2}|}=-\sum_{i=1}^{N+1} e_i.
\end{align}
Substituting \eqref{new_alpha} into \eqref{close_loop} yields 
\begin{align}
\label{closed_loop_1}
\dot{\bold{e}}=(A+C_0)\bold{e}
\end{align}
where the switching matrix $C_0\in\mathbb{R}^{(2N+1)\times (2N+1)}$ is given by
\begin{tikzpicture}[overlay, remember picture]
     \node[anchor=west] at (5.3, -2.20) {$\scriptstyle (N+2)-th$};
     \node[anchor=west] at (5.3, -1.90) {$\scriptstyle (N+1)-th$};
     \node[anchor=west] at (1.1, -0.3) {\rotatebox{270}{$\left\{ \begin{array}{c} \\ \\ \\  \\ \\  \end{array} \right.$}};
     \node[anchor=west] at (1.9, 0.20) {$\scriptstyle n-1$};
\end{tikzpicture}
\begin{align}
\label{matrix_C_0}
C_0={\scriptsize
\begin{bmatrix}
-k_2 & \cdots & -k_2 & 0 & \cdots & 0\\
0 & \cdots & 0 & 0 & \cdots & 0\\
\vdots & \vdots & \vdots & \ddots & \vdots\\
0 & \cdots & 0 & 0 & \cdots & 0\\
\hline
-k_2 & \cdots & -k_2 & 0 & \cdots & 0\\
\hline
k_2 & \cdots & k_2 & 0 & \cdots & 0\\
\hline
0 & \cdots & 0 & 0 & \cdots & 0\\
\vdots & \vdots & \vdots & \ddots & \vdots\\
0 & \cdots & 0 & 0 & \cdots & 0\\
k_2 & \cdots & k_2 & 0 & \cdots & 0\\
\end{bmatrix}}~~~~~~~~~~~~~~.
\end{align}
As either vehicle $j$ or $k$ passes through the intersection according to the switching sequence in \eqref{cycle_sequence}, the corresponding switching matrices $C_i$ can be generated from $C_0$ through circular permutations as
\begin{align}
\label{switching_matrix_property}
C_i=P^{iN}C_0P^{-iN}\in\mathbb{R}^{(2N+1)\times (2N+1)}, \qquad \forall i\in\mathcal S,
\end{align}
where the index set $\mathcal{S}=\{0,1,\cdots,2N\}$ and the circular permutation matrix $P$ is given by
\begin{align}
\label{matrix_P}
P=&{\scriptsize
\begin{bmatrix}
 0 & 0 &\cdots & 0 & 1\\
 1 & 0 &\cdots & 0 & 0\\
 0 & 1 &\cdots & 0 & 0\\
 \vdots & \vdots & \ddots & \vdots & \vdots\\
 0 & 0 &\cdots & 1 & 0\\
\end{bmatrix}\scriptsize}\in\mathbb{R}^{(2N+1)\times (2N+1)}.
\end{align}
Therefore, under the switching sequence in \eqref{cycle_sequence}, the error dynamics in \eqref{close_loop} can be represented as a switched system depending only on $\bold{e}$ below,
\begin{align}
\label{closed_loop_2}
\dot{\bold{e}}=(A+C_{\sigma(t)})\bold{e},
\end{align} 
where the matrix $A$ is given in \eqref{error_matrix}, $C_{\sigma(t)}\in\mathbb{R}^{(2N+1)\times(2N+1)}$ is the switching matrix, and $\sigma(t):[0, +\infty)\rightarrow \mathcal{S}$ is a switching signal taking values in the index set $\mathcal{S}$ in \eqref{switching_matrix_property}. The following lemma establishes two key properties of this switched system~\eqref{closed_loop_2}.

\begin{lemma}
\label{lemma_invariant}
Consider the switched system~\eqref{closed_loop_2}  governed by the finite switching sequence in \eqref{cycle_sequence} satisfies the properties:
\begin{enumerate}
\item For all possible system matrices $A+C_{\sigma(t)}$, the eigenvalues are identical.

\item The state $\bold{e}(t)$ remains in the invariant subspace
$\bold{1}^{\perp}=\{\bold{e}\mid \bold{1}_{2N+1}^\top \bold{e}=0\}$ for all $ t>0$.
\end{enumerate}
\end{lemma}

\begin{proof}
Based on \eqref{matrix_P}, the cyclic matrix $A$ in~\eqref{error_matrix} satisfies
\begin{align}
\label{cyclic_matrix_property}
A=P^{j}AP^{-j}, \qquad \forall j\in\mathbb{Z}^+.
\end{align}
Combining \eqref{switching_matrix_property} and \eqref{cyclic_matrix_property} together yields 
\begin{align}
A+P^{iN}C_0P^{-iN}=P^{iN}(A+C_0)P^{-iN}, \qquad \forall i\in\mathcal S,
\end{align}
which implies that all possible dynamic matrices $P^{iN}(A+C_0)P^{-iN}$ in \eqref{closed_loop_2} satisfy the similarity transformation, and have the same eigenvalues.
Then, the proof of Claim 1) in Lemma~\ref{lemma_invariant} is thus completed.

Together with $\bold{e}$ in \eqref{relative_angle}, one has that the states $\bold{e}(t)$ stay in the invariant subspace 
\begin{align}
\label{subspace}
\bold{1}^{\perp}=\{\bold{e}~| ~\bold{1}_{2N+1}\t\bold{e}=0\},
\end{align}
which implies that $\bold{1}_{2N+1}\t\bold{e}(t)=0$ for all $t>0$. The proof of Claim 2) is thus completed.
\end{proof}

\begin{remark}
The reason for selecting the first and $(N+2)$-th vehicles (i.e., $j=1$ and $k=N+2$ in \eqref{initial_intersection}) to define the initial value $d_{1,N+2}$ is that this choice leads to a concise expression of $d_{1,N+2}$ in terms of $\sum_{i=1}^{N} e_i$, which facilitates the formulation of the switching matrix $C_i$ in \eqref{switching_matrix_property}. For other scenarios in which different vehicles meet at the intersection in Fig.~\ref{vehicle_frame} (a), the vehicles can be renumbered such that they correspond to the first and $(N+2)$-th vehicles as well.
\end{remark}


\subsection{The Convergence of the Platoon}
\label{sub_platoon_convergence}
We prove platoon convergence in {\bf P1}. The analysis relies on the Euclidean matrix~measure in the following lemma. 

\begin{lemma} \label{lemma_matrix_meature}
\cite{khalil2002nonlinear}: Consider a linear system $\dot{\boldsymbol{x}} = H \boldsymbol{x}$. If there exists a constant $c>0$ such that $\mu_2(H) \le -c < 0$, where
\begin{equation}
\label{mu_matrix}
\mu_2(H)=\lambda_{\max}\!\left(\frac{H+H^\top}{2}\right),
\end{equation}
then $\|\dot{\bold{x}}(t)\| \le \mu_2(H)\|\bold{x}(t)\| \le -c\,\|\bold{x}(t)\|$, which implies that $\dot{\bold{x}}=H\bold{x}$ is exponentially stable, i.e., $\|\boldsymbol{x}(t)\|\le e^{-ct}\|\boldsymbol{x}(0)\|$ for all $t>0$.
\end{lemma}

Since Lemma~\ref{lemma_invariant} shows that the switched systems \eqref{closed_loop_2} share the same eigenvalues, Lemma~\ref{lemma_matrix_meature} provides a convenient tool for relating platoon convergence to the effects of the switching signal $\sigma(t)$. It remains to show that the reduced subsystem matrices admit a common negative matrix-measure bound.










\begin{lemma}
\label{lemma_platoon_convergence}
Under Assumptions~\ref{assump_two_vehicle_intersection}-\ref{initial_error}, all the vehicles $\mathcal V$ governed by the controller \eqref{vehicle_controller} and the switching sequence in~\eqref{cycle_sequence} can achieve the exponential convergence of the platoon, i.e., $\lim_{t\rightarrow\infty}\bold{e}(t)=\bold{0}_{2N+1}$ in {\bf P1} of Definition~\ref{def_platoon_avoidance}, if the parameters $k_1, k_2$ satisfy the condition \eqref{convergence_parameter_k} in Theorem~\ref{theorem_main}.
\end{lemma}

\begin{proof}
Recalling the invariant subspace $\mathbf{1}^\perp$ in \eqref{subspace}, we define $T\in\mathbb{R}^{(2N+1)\times 2N}$ to be a matrix whose columns form an orthonormal basis of $\mathbf{1}^\perp$, i.e.
\begin{equation}
\label{orthogonal_basis}
T\t T =I_{2N},\ \mathbf{1}_{2N+1}\t T = \bold{0}_{2N}. 
\end{equation}
Since the error state $\bold{e}$ satisfies $\bold{1}_{2N+1}\t \bold{e}(t)=0$ for all $t>0$ in \eqref{subspace}, one has that $\bold{e}(t)\in\bold{1}^{\perp}$, which implies that the error vector can be uniquely written as 
\begin{equation}
\label{reduced_state}
\bold{e} = T \bold{z}, 
\end{equation} 
where $\bold{z}\in\mathbb{R}^{2N}$ is the reduced coordinate. Multiplying both sides in \eqref{reduced_state} by $T\t$ and substituting \eqref{orthogonal_basis} gives 
\begin{equation}
\label{reduced_z}
\bold{z} = T\t \bold{e} . 
\end{equation}
By taking the derivative of $\bold{z}$ in \eqref{reduced_z} and substituting \eqref{closed_loop_2} and~\eqref{reduced_state}, one has that
\begin{align}
\dot{\bold{z}} = T\t (A+C_{\sigma(t)})T \bold{z}.
\end{align}
Let $D_{\sigma(t)} = T\t (A+C_{\sigma(t)})T$
be the reduced subsystem matrix. Then, the error dynamics $\bold{e}$ in \eqref{closed_loop_2} reduce to the $(n-1)$-dimensional switched systems $\bold{z}$ below
\begin{equation} 
\label{redcued_switched_system}
\dot{\bold{z}} = D_{\sigma(t)} \bold{z}. 
\end{equation} 
It follows from \eqref{redcued_switched_system} that the eigenvalues of $D_{\sigma(t)}$ coincide with the nonzero $2N$ eigenvalues of $(A + C_{\sigma(t)})$. This implies that the convergence of the error system \eqref{closed_loop_2} is equivalent to the stability of the reduced switched subsystem in \eqref{redcued_switched_system}, i.e.,
\begin{align}
\label{equivalent_two_system}
\lim_{t\rightarrow\infty}\bold{z}=\bold{0}_{2N}\Longrightarrow\lim_{t\rightarrow\infty}\bold{e}=\bold{0}_{2N+1}
\end{align}
with \eqref{reduced_state}. Meanwhile, together with Claim 1) in Lemma~\ref{lemma_invariant}, it follows that all possible matrices $D_i$, $i \in \mathcal{S}$, share the same eigenvalues regardless of the switching signal $\sigma(t)$.

Accordingly, we can calculate the common Euclidean matrix measure of all possible matrices $D_i$ to prove the convergence of $\bold{z}$. In particular, for arbitrary $i\in\mathcal{S}$, it follows from~\eqref{redcued_switched_system} that $D_i = T\t A T + T\t C_i T$,
which implies that
\begin{align}
\label{matrix_decoupling}
\frac{D_i+D_i\t}{2}=\frac{T\t AT + T\t A\t T}{2} + \frac{T\t C_iT + T\t C_i\t T}{2}.
\end{align}
Since the matrices $(D_i+D_i\t), (T\t AT + T\t A\t T), (T\t C_iT + T\t C_i\t T)$ are all symmetric, it follows from \eqref{mu_matrix} and \eqref{matrix_decoupling} that

\begin{equation}
\label{matrix_measure_inequality}
\mu_2(D_i)
\le
\lambda_{\max}\!\left(\frac{A+A^T}{2}\Big|_{\mathbf{1}^\perp}\right)
+
\lambda_{\max}\!\left(\frac{C_i+C_i^T}{2}\right).
\end{equation}
To calculate the upper bound of $\mu_2(D_i)$ in \eqref{matrix_measure_inequality}, we analyze the symmetric parts of $A$ and $C_i$ for all $i\in\mathcal S$, separately.





\textbf{Step 1: Euclidean matrix measure of $A$.}
Recall that the matrix $A$ in \eqref{error_matrix} is rewritten to be
$A = k_1(P-I_{2N+1})$,
where the cyclic permutation matrix $P$ is given in \eqref{matrix_P}. Then, the symmetric part of $A$ is
\begin{equation}
\label{symmetric_A}
\frac{A+A^T}{2}
=
\frac{k_1}{2}(P+P^T-2I_{2N+1}).
\end{equation}
According to the special structure of $P$ in \eqref{matrix_P}, one has that its eigenvalues are
\begin{equation}
\omega_m = e^{i\frac{2\pi m}{2N+1}}, \qquad m\in\mathbb{Z}_0^{2N},
\end{equation}
which implies that the eigenvalues of $P+P^T$ are $\omega_m + \omega_m^{-1}
=
2\cos\frac{2\pi m}{2N+1}.$
It follows from \eqref{symmetric_A} that all the possible eigenvalues of $(A+A\t)/2$ becomes
\begin{equation}
\label{matrix_measure_A}
\lambda_m\left(\frac{A+A\t}{2}\right)
=
k_1\left(\cos\frac{2\pi m}{2N+1}-1\right),
\qquad m\in\mathbb{Z}_0^{2N}.
\end{equation}
Since $\cos(0)=1$, the largest eigenvalue in \eqref{matrix_measure_A} is $0$ corresponding to the vector
$\mathbf{1}$. However, when restricting the dynamics $\bold{z}$ in \eqref{redcued_switched_system} to the subspace
$\mathbf{1}^\perp$, the mode $m=0$ is removed, and the largest remaining eigenvalue of $(A+A\t)/2$ is attained at $m=1$, i.e., 
\begin{equation}
\lambda_{\max}\!\left(\frac{A+A^T}{2}\Big|_{\mathbf{1}^\perp}\right)
=
k_1\left(\cos\frac{2\pi}{2N+1}-1\right).
\end{equation}

\textbf{Step 2: Euclidean matrix measure of $C_i$ for all $i\in\mathcal S$.}
Since the matrix $C_i$ satisfies $C_i=P^{iN}C_0P^{-iN}$ in \eqref{switching_matrix_property}, one has that all matrices $C_i$ are similar and share the same eigenvalues. Thus, we only need to calculate the matrix measure of $C_0$ in \eqref{matrix_C_0} to represent $C_i$ for all $i\in\mathcal S$. To proceed, we first define the basis vectors $\bold{a}, \bold{b}$ below
\begin{align}
\label{vector_a_b}
\bold{a}=&
\begin{bmatrix}
-1 & 0 & \cdots & 0 & -1 & 1 & 0 & \cdots & 0 & 1 
\end{bmatrix}\in\mathbb{R}^{2N+1},\nonumber\\
\bold{b}=&
\begin{bmatrix}
1 & \cdots & 1 & 0 & \cdots & 0
\end{bmatrix}\in\mathbb{R}^{2N+1},
\end{align}
one has that the matrix $C_0$ in \eqref{matrix_C_0} can be rewritten as a rank-one matrix $C_0 = k_2 \bold{a} \bold{b}^T$,
where the vectors $\bold{a},\bold{b}$ in \eqref{vector_a_b} satisfy
\begin{equation}
\label{vector_property}
\|\bold{a}\|=2, \|\bold{b}\|=\sqrt{N+1}, \bold{a}^T \bold{b}=\bold{b}^T \bold{a} = -2.
\end{equation}
Then, the symmetric part of $C_0$ becomes
\begin{equation}
\label{sysmetric_C_0}
\frac{C_0 + C_0^T}{2}
=
\frac{k_2}{2}(\bold{a}\bold{b}^T + \bold{b}\bold{a}^T).
\end{equation}
Based on \eqref{sysmetric_C_0}, we can further define the matrix $M$ and the corresponding eigenvector $\nu$ to be 
\begin{align*}
M=\frac{\bold{a}\bold{b}^T + \bold{b}\bold{a}^T}{2}, 
\nu=c_1\bold{a}+c_2\bold{b},
\end{align*}
with arbitrary parameters $c_1, c_2$, which implies that 
\begin{align}
\label{M_equality}
M\nu=&\frac{1}{2}(\bold{a}\bold{b}^T\nu + \bold{b}\bold{a}^T\nu)\nonumber\\
=&\frac{1}{2}\Big(\bold{a}(c_1\bold{b}^T\bold{a}+c_2\bold{b}^T\bold{b}) + \bold{b}(c_1\bold{a}^T\bold{a}+c_2\bold{a}^T\bold{b})\Big).
\end{align}
Substituting \eqref{vector_property} into \eqref{M_equality} yields
\begin{align}
\label{M_equality_1}
M\nu=
&\left(-c_1+\frac{N+1}{2}c_2\right)\bold{a}+(2c_1-c_2)\bold{b}.
\end{align}
Since $M\nu=\lambda \nu=\lambda (c_1\bold{a}+c_2\bold{b})$ with $\lambda$ being the corresponding eigenvalue, it follows from \eqref{M_equality_1} that 
\begin{align*}
&-c_1+\frac{N+1}{2}c_2=\lambda c_1,~2c_1-c_2=\lambda c_2.
\end{align*}
These two equations can be rewritten into the following compact form 
\begin{align}
\label{parameter_equality_matrix}
\begin{bmatrix}
-1-\lambda & \frac{N+1}{2}\\
2 & -1-\lambda
\end{bmatrix}
\begin{bmatrix}
c_1\\
c_2
\end{bmatrix}
=\bold{0}_2.
\end{align}
To ensure that 
\eqref{parameter_equality_matrix} holds for arbitrary $(c_1, c_2)\neq(0,0)$, one has that 
\begin{equation}
\mathrm{det}\bigg(\begin{bmatrix}
-1-\lambda & \frac{N+1}{2}\\
2 & -1-\lambda
\end{bmatrix}\bigg)=0,
\end{equation}
which implies that $\lambda = -1 \pm \sqrt{N+1}$.
Therefore, the largest eigenvalue of the symmetric part of $C_i$ is
\begin{align}
\label{matrix_measure_C}
\lambda_{\max}\!\left(\frac{C_i+C_i^T}{2}\right)=&\lambda_{\max}\!\left(\frac{C_0+C_0^T}{2}\right) \nonumber\\
=& k_2(\sqrt{N+1}-1) \qquad \forall i\in\mathcal S.
\end{align}

\textbf{Step 3: The convergence of $\bold{e}(t)$ in \eqref{closed_loop_2}.}
By combining \eqref{matrix_measure_A} and \eqref{matrix_measure_C} together, it follows from \eqref{matrix_measure_inequality} that the upper bound of $\mu_2(D_i)$ becomes
\begin{equation}
\label{matrix_measure_inequality_1}
\mu_2(D_i)
\le
k_1\left(\cos\frac{2\pi}{2N+1}-1\right)
+
k_2(\sqrt{N+1}-1).
\end{equation}
Substituting the condition of  \eqref{convergence_parameter_k} into \eqref{matrix_measure_inequality_1} yields $\mu_2(D_i)
\le -c$ for all $i\in\mathcal S$,
which follows from \eqref{redcued_switched_system} that 
\begin{equation} 
\label{redcued_switched_system_1}
\dot{\bold{z}}(t) = D_{\sigma(t)} \bold{z}(t)\leq -c\|\bold{z}(t)\|. 
\end{equation} 
Using Lemma~\ref{lemma_matrix_meature} and 
\eqref{redcued_switched_system_1}, 
we conclude that the reduced switched subsystem \eqref{redcued_switched_system} is uniformly exponentially stable, i.e., $\|\bold{z}(t)\| \le e^{-ct}\|\bold{z}(0)\|\Longrightarrow\lim_{t\rightarrow\infty}\|\bold{z}(t)\|=0$,
which follows from \eqref{equivalent_two_system} that $\lim_{t\rightarrow\infty}\bold{e}=\bold{0}_{2N+1}$, exponentially. The proof is thus completed.
\end{proof}




\subsection{Safe- and Smooth-Crossing Platoon}
\label{sub_smooth_crossing}
\begin{lemma}
\label{lemma_smooth_crossing}
Under Assumptions~\ref{assump_two_vehicle_intersection}-\ref{initial_error} and all hypotheses in Lemma~\ref{lemma_platoon_convergence}, all vehicles $\mathcal V$ governed by the controller \eqref{vehicle_controller} and the switching sequence in~\eqref{cycle_sequence} can guarantee the smooth crossing at the intersection, i.e., $\lim_{t\rightarrow\infty}\dot{\theta}_i(t)=\lim_{t\rightarrow\infty}\dot{\theta}_l(t)=v_d$ for all $ i\neq l\in\mathcal V,$ in {\bf P3} of Definition~\ref{def_platoon_avoidance}.
\end{lemma}

\begin{proof}
Recalling the condition of $\lim_{t\rightarrow\infty}\bold{e}=\bold{0}_{2N+1}$ in Lemma~\ref{lemma_platoon_convergence}, one has that $\lim_{t\rightarrow\infty}\dot{\bold{e}}=\bold{0}_{2N+1}$, which follows from \eqref{phase_dynamics} and \eqref{relative_angle} that
\begin{align}
\label{convergence_dot_theta}
&\lim_{t\rightarrow\infty}\dot{\theta}_{i+1}(t)-\dot{\theta}_{i}(t)=0,\Longleftrightarrow
\lim_{t\rightarrow\infty}u_{i+1}(t)-u_{i}(t)=0\nonumber\\&(i+1=1, \mathrm{if}~i=2N+1), \forall i\in\mathcal V. 
\end{align}
Meanwhile, recalling the detailed controller $u_i$ in \eqref{vehicle_controller}, it follows from \eqref{convergence_dot_theta} that $\lim_{t\rightarrow\infty}\eta_{i+1}(t)-\eta_{i}(t)=0$ for all $i\in\mathcal V$.
Since $\eta_i$ in \eqref{potential_function} might be nonzero only if $i=j$ or $i=k$ at the intersection, one has that all the additional collision avoidance $\eta_i$ terms converge to zeros, i.e., $\lim_{t\rightarrow\infty}\eta_i(t)=0$ for all $i\in\mathcal V$.
By combining \eqref{convergence_dot_theta}, and the condition of $\lim_{t\rightarrow\infty}\bold{e}=\bold{0}_{2N+1}$ in Lemma~\ref{lemma_platoon_convergence} together, one has that $\lim_{t\rightarrow\infty}\dot{\theta}_i(t)=\lim_{t\rightarrow\infty}\dot{\theta}_l(t)=\lim_{t\rightarrow\infty}u_i(t)=\lim_{t\rightarrow\infty}u_l(t)=v_d$.
\end{proof}


\begin{lemma}
\label{lemma_collision_avoidance_vehicle}
Under Assumptions~\ref{assump_two_vehicle_initial_position}-\ref{initial_error} and all hypotheses in Lemma~\ref{lemma_platoon_convergence}, all vehicles $\mathcal V$ governed by the controller \eqref{vehicle_controller} and the switching sequence in~\eqref{cycle_sequence} can guarantee the collision avoidance between adjacent vehicles, i.e., $\theta_{i+1}(t)-\theta_i(t)+\xi_i> 0$ for all $ i\in\mathcal V,$ in {\bf P2} of Definition~\ref{def_platoon_avoidance}.
\end{lemma}

\begin{proof}

From Lemma~\ref{lemma_platoon_convergence}, the reduced switched subsystem is uniformly exponentially stable. In particular, it follows from \eqref{redcued_switched_system_1} that
$\|\bold{z}(t)\|\leq e^{-ct}\|\bold{z}(0)\|$ for all $t\geq 0$.
Since $\bold{e}=T\bold{z}$ in \eqref{reduced_state} and the columns of $T$ form an orthonormal basis of $\bold{1}^\perp$, one has that $\|\bold{e}(t)\|=\|\bold{z}(t)\|\leq \|\bold{e}(0)\|$. Moreover, Assumption~\ref{initial_error} implies that $
\|\bold{e}(0)\|\leq \sqrt{2N+1}\epsilon<{2\pi}({2N+1}).$
Thus, for each $i\in\mathcal V$, one has that
\[
e_i(t)\geq -|e_i(t)|\geq -\|\bold{e}(t)\|>-\frac{2\pi}{2N+1}, \qquad t>0.
\]
By the definition of $e_i$ in \eqref{relative_angle}, we then obtain $\theta_{i+1}(t)-\theta_i(t)+\xi_i(t)=e_i(t)+{2\pi}/({2N+1})>0$. The proof is completed.
\end{proof}

\begin{proof}[Proof of Theorem \ref{theorem_main}] The proof follows directly from the application of 
Lemmas~\ref{lemma_existence}-\ref{lemma_collision_avoidance_vehicle}.
\end{proof}

\begin{figure}[!htb]
\centering
\includegraphics[width=7.5cm]{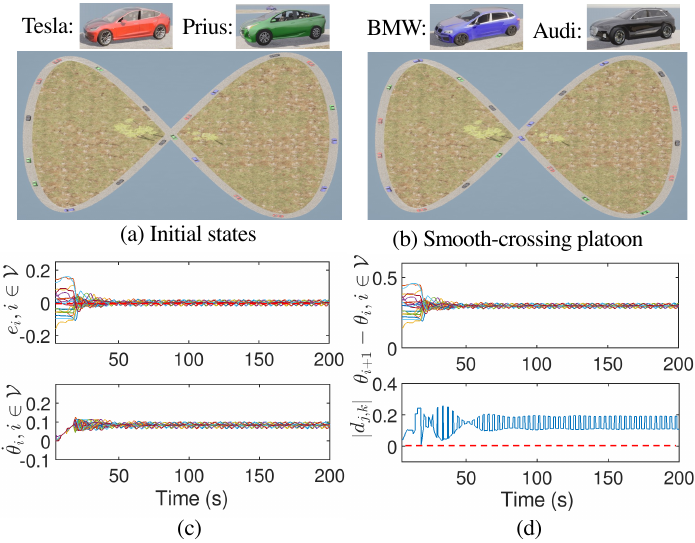}
\vspace{-0.4em}
\caption{ {\it Safe- and smooth-crossing platoon} on figure-eight path in CARLA.
(a) Initial states. (b) Safe- and smooth-crossing platoon. (c) Error $e_i$ and phase velocity $\dot{\theta}_i$. (d) Adjacent phase error $\theta_{i+1}-\theta_i$ and switching distance $|d_{j,k}|$ at the intersection.}
\label{path_eight}
\end{figure}
\begin{figure}[!htb]
\centering
\includegraphics[width=7.2cm]{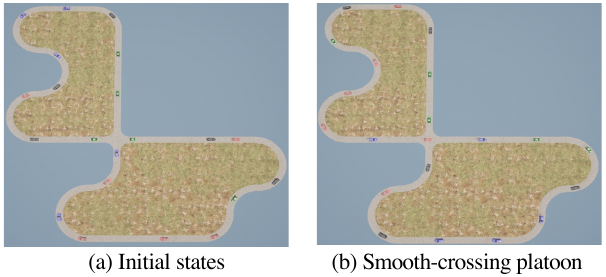}
\vspace{-0.4em}
\caption{Validation on random urban paths with an intersection in CARLA.
(a) Initial states. (b) Safe- and smooth-crossing platoon. }
\label{path_random}
\end{figure}

\section{Simulations}
\label{sec_simulation}

We validate the proposed algorithm in the high-fidelity CARLA simulator \cite{dosovitskiy2017carla} using $21$ vehicles on two paths, with $k_1=3.5$, $k_2=25$, and $R=\pi/43$. For the figure-eight path in Fig.~\ref{path_eight}, the vehicles start near the desired platoon configuration and converge to the desired {\it safe-} and {\it smooth-crossing platoon}. Fig.~\ref{path_eight}(c) shows that $e_i$ converges to zero and $\dot{\theta}_i$ converges to a constant, which verifies platoon convergence and smooth crossing ({\bf P1} and {\bf P3}). Fig.~\ref{path_eight}(d) shows that $\theta_{i+1}(t)-\theta_i(t)>0$ and $|d_{j,k}(t)|>0$, which verifies adjacent-vehicle and intersection collision avoidance ({\bf P2} and {\bf P4}). The oscillation and strict positivity of $|d_{j,k}|$ show repeated but collision-free intersection crossings. Fig.~\ref{path_random} demonstrates adaptability on a random urban path. More details are provided in the online video at \href{https://youtu.be/VCs-RUui6oQ}{https://youtu.be/VCs-RUui6oQ}.

\section{Conclusion}
\label{sec_conclusion}

This paper designed a distributed controller for {\it safe-} and {\it smooth-crossing platoons} at an unsignalized intersection. We proved local exponential convergence of the {\it switched closed-loop system} on a closed path with a single intersection. Future work will investigate the scalability of the analysis to uneven and multiple intersections, as well as open paths. 

\appendices

\bibliographystyle{IEEEtran}
\bibliography{IEEEabrv,ref}

\end{document}